\documentclass[aps,prl,reprint,amsmath,amssymb,superscriptaddress]{revtex4-2}
\usepackage{blindtext}
\usepackage{graphicx}% Include figure files
\usepackage{dcolumn}% Align table columns on decimal point
\usepackage[colorlinks,urlcolor=blue,linkcolor=blue,citecolor=blue]{hyperref}
\usepackage{bm}% bold math
\usepackage{subfigure}
\graphicspath{{./Pictures/}}
\usepackage{xcolor}
\usepackage{lipsum}
\usepackage{xcolor}

\begin{document}
\title{Determination of Multi-mode Motional Quantum States in a Trapped Ion System}
\author{Zhubing Jia}
\email{zhubing.jia@duke.edu}
\affiliation{Duke Quantum Center, Duke University, Durham, NC 27701, USA}
\affiliation{Department of Physics, Duke University, Durham, NC 27708, USA}
\author{Ye Wang}
\email{ye.wang2@duke.edu}
\affiliation{Duke Quantum Center, Duke University, Durham, NC 27701, USA}
\affiliation{Department of Electrical and Computer Engineering, Duke University, Durham, NC 27708, USA}
\author{Bichen Zhang}
\altaffiliation[Present address: ]{Department of Electrical and Computer Engineering, Princeton University, Princeton, NJ 08544, USA}
\affiliation{Duke Quantum Center, Duke University, Durham, NC 27701, USA}
\affiliation{Department of Electrical and Computer Engineering, Duke University, Durham, NC 27708, USA}
\author{Jacob Whitlow}
\affiliation{Duke Quantum Center, Duke University, Durham, NC 27701, USA}
\affiliation{Department of Electrical and Computer Engineering, Duke University, Durham, NC 27708, USA}
\author{Chao Fang}
\affiliation{Duke Quantum Center, Duke University, Durham, NC 27701, USA}
\affiliation{Department of Electrical and Computer Engineering, Duke University, Durham, NC 27708, USA}
\author{Jungsang Kim}
\affiliation{Duke Quantum Center, Duke University, Durham, NC 27701, USA}
\affiliation{Department of Physics, Duke University, Durham, NC 27708, USA}
\affiliation{Department of Electrical and Computer Engineering, Duke University, Durham, NC 27708, USA}
\affiliation{IonQ, Inc., College Park, MD 20740, USA}
\author{Kenneth R. Brown}
\email{ken.brown@duke.edu}
\affiliation{Duke Quantum Center, Duke University, Durham, NC 27701, USA}
\affiliation{Department of Physics, Duke University, Durham, NC 27708, USA}
\affiliation{Department of Electrical and Computer Engineering, Duke University, Durham, NC 27708, USA}
\affiliation{Department of Chemistry, Duke University, Durham, NC 27708, USA}

\date{\today}

\begin{abstract}
Trapped atomic ions are a versatile platform for studying interactions between spins and bosons by coupling the internal states of the ions to their motion. Measurement of complex motional states with multiple modes is challenging, because all motional state populations can only be measured indirectly through the spin state of ions. Here we present a general method to determine the Fock state distributions and to reconstruct the density matrix of an arbitrary multi-mode motional state. We experimentally verify the method using different entangled states of multiple radial modes in a 5-ion chain. This method can be extended to any system with Jaynes-Cummings type interactions.
\end{abstract}

\maketitle

{\textit{Introduction.---}}Trapped atomic ion systems are a flexible platform for quantum simulations. High-fidelity single-qubit rotations and two-qubit entangling gates have been realized in trapped ion systems~\cite{Ballance2016, Gaebler2016, Wang2020, brown2021, Baldwin2021, srinivas2021}, enabling spin-based digital quantum simulations~\cite{Landsman2019, nam2020ground, Monroe2021, Lanyon2011, Qsim2021}. Apart from digital simulations, the Jaynes-Cummings type interactions between ion spins and motional phonons of the harmonic potential offer a natural platform for analog quantum simulations of spin-boson coupling~\cite{Porras2012, toyoda2015, Gorman2018, zhang2018, Lv2018, mei2021experimental, cai2021observation}. 
We can also take  advantage of combining discrete and continuous variables for quantum computation based on encoding qubits in bosonic modes~\cite{Gottesman2001, fluhmann2019encoding} and demonstrating hybrid quantum computation~\cite{ortiz2017continuous, Gan2020Hybrid}. These applications require coherent manipulation and measurement of complex bosonic states, and specifically in trapped-ion systems, motional states with potentially multiple modes.

Comparing with qubit spin state which in trapped-ion systems can be measured with less than 0.1\% error~\cite{Olmschenk2007, noek2013high, Harty2014, Myerson2008, Todaro2021}, the motional Fock state distributions, i.e., the probabilities of the state in each motional Fock state basis, cannot be measured directly: the states need to be mapped onto qubit spin states. For single-mode motional states, by driving Jaynes-Cummings type interactions, the Fock state distributions have been characterized~\cite{Meekhof1996} and the density matrix as well as Wigner functions of non-classical motional states have been reconstructed~\cite{Leibfried1996, Kienzler2016, um2016phonon, Lv2017}. With multiple motional modes, one cannot simply measure each mode and then combine the results, since the entanglement between different modes will be traced out. Several attempts have been made to resolve two-mode Fock state distributions and to verify certain types of entangled two-mode motional states, but these approaches either introduce overhead from phonon arithmetic operations and multiple rounds of detection or are not available for normal modes~\cite{Zhang2018NOON, Ohira2019}. 
Methods for measuring many-mode Fock states remain under explored.

Here we propose a general method to efficiently determine the Fock state distributions of an arbitrary multi-mode motional state in a trapped-ion system by applying coherent manipulations and joint qubit state measurement on multiple ions, and experimentally demonstrate the measurement of the Fock state distributions of different two-mode and three-mode motional states. We discuss the theory of reconstructing the density matrix of a $d$-mode motional quantum state by extending the single-mode result in Ref.~\cite{Leibfried1996}, and reconstruct the density matrix of a 2-mode motional Bell state. 
This measurement requires Jaynes-Cummings type interactions and individual measurements on n-spins, and is a useful tool for studying the behavior of multiple bosonic modes in any system that meets these requirements.

{\textit{Theory.---}}We briefly describe the effective Hamiltonians that drive the coherent manipulations between the spin states $\left|\uparrow\right>$ and $\left|\downarrow\right>$ and motional Fock states $\left|n\right>$ of ions (for a detailed derivation of these Hamiltonians see Ref. \cite{Wineland1998}).
In the Lamb-Dicke limit where the motion of the ion is small compared to the laser wavelength, we can tune the lasers to  drive (anti-)Jaynes-Cummings transitions  $\left|\downarrow\right>\left|n\right>\leftrightarrow\left|\uparrow\right>\left|n\pm 1\right>$, namely blue sideband (BSB) and red sideband (RSB) transitions:
\begin{align}
    \begin{split}
        H_{BSB} &= i\Omega_b\left(\sigma_+a^\dagger e^{i\phi_b}-\sigma_-a e^{-i\phi_b}\right)\\
        H_{RSB} &= i\Omega_r\left(\sigma_+a e^{i\phi_r}-\sigma_-a^\dagger e^{-i\phi_r}\right)
    \end{split}
\end{align}
where $\sigma_{+(-)}$ are spin raising(lowering) operators, $a^\dag(a)$ are motional raising(lowering) operators, and $\Omega_{r(b)}$ and $\phi_{r(b)}$ are the Rabi frequency and phase of R(B)SB transitions. By applying RSB and BSB transitions simultaneously with the same Rabi frequency $\Omega$, one can achieve a spin-state dependent coherent displacement Hamiltonian~\cite{Monroe1996}:
\begin{align}
    H_{D} &= i\Omega\left(\sigma_+a e^{i\phi_r}-\sigma_-a^\dagger e^{-i\phi_r}+\sigma_+a^\dagger e^{i\phi_b}-\sigma_-a e^{-i\phi_b}\right)\nonumber\\
    &= i\Omega\left(\sigma_+e^{i\phi_s}-\sigma_-e^{-i\phi_s}\right)\left(a^\dagger e^{i\phi_m}+ae^{-i\phi_m}\right)
    \label{disp}
\end{align}
where $\phi_{s} = \left(\phi_b+\phi_r\right)/2$ and $\phi_{m} = \left(\phi_b-\phi_r\right)/2$. When the spin state of the ion is the +1-eigenstate of $i\left(\sigma_+e^{i\phi_s}-\sigma_-e^{-i\phi_s}\right)$, the motional state experiences a displacement operation $D\left(\Omega te^{i\phi_m}\right)$, where $t$ is the total displacement time. We can apply the operations above to multiple motional modes and create various non-classical multi-mode motional states.

After preparing a $d$-mode motional target state $\rho$, we can measure the Fock state distributions of the state, $P_{k_1,\cdots, k_d}=\left<k_1\cdots k_d\right|\rho\left|k_1\cdots k_d\right>$. Here $\left|k_1\cdots k_d\right>$ denotes the $d$-mode Fock state basis with $k_j$ phonons in the $j$th mode. Previous works have obtained 2-mode Fock state probabilities, but % was only verified with specific target states
this was with the cost of serials of composite pulses and projection measurements~\cite{Zhang2018NOON}. Based on our ability to do individual gate operations and detection on each ion, we develop a faster and more straightforward way to determine the values of $P_{k_1, \cdots, k_d}$. After the target state is prepared and $d$ ions in the chain are set to $\left|\downarrow\right>$ state, we drive BSB transitions of the $d$ modes on $d$ ions ($d$-mode BSB) separately with different Rabi frequencies 
\begin{align}
    H_{d\text{-mode~BSB}} = \sum_{j=1}^{d}i\Omega_j\left(\sigma_{+,j}a_j^\dagger e^{i\phi_j}-\sigma_{-,j}a_j e^{-i\phi_j}\right)
\end{align}
for the same amount of time $t$. The probability of measuring all ions in $\left|\downarrow\right>$ state  $\mathcal{P}_{\downarrow\cdots\downarrow}\left(t\right)$ has the form

\begin{align}
    \mathcal{P}_{\downarrow\cdots\downarrow}\left(t\right) = \sum_{k_1,\cdots, k_d=0}^{\infty}&P_{k_1,\cdots, k_d}\prod_{j=1}^{d}\cos^2\left(\sqrt{k_j+1}\Omega_{j}t\right)
    \label{dDBSB}
\end{align}
Then the values of $P_{k_1,\cdots,k_d}$ can be fit from this joint spin state distribution by using $\prod_{j=1}^{d}\cos^2\left(\sqrt{k_j+1}\Omega_{j}t\right)$ as a basis set. With motional decoherence, Eq.~(\ref{dDBSB}) works as an appropriate approximation when the motional coherence time $\tau \gg t$. Similarly, the populations of other joint spin state configurations can also be used for fitting $P_{k_1,\cdots,k_d}$ with different sets of basis.

With the ability to fit all Fock state probabilities of a density matrix, we can further determine the off-diagonal terms of the target state density matrix. As in Ref.~\cite{Leibfried1996}, we apply displacement operations $D\left(\alpha\right)=\exp\left(\alpha a^\dagger-\alpha^*a\right)$ on each motional mode with certain displacement amplitudes and phases, and then measure the Fock state distributions of the displaced target state $Q_{k_1\cdots k_d}\left(\alpha_1,\cdots, \alpha_d\right)=\left<k_1\cdots k_d\right|\prod_{j=1}^d D_j^\dagger\left(\alpha_j\right)
% D_2^{\dagger}\left(\alpha_2\right)D_1^{\dagger}\left(\alpha_1\right)
\rho 
\prod_{j=1}^d D_j\left(\alpha_j\right)
\left|k_1\cdots k_d\right>$. Here $D_j\left(\alpha_j\right)$ represents displacement operations on mode $j$ with displacement amount $\alpha_j$. Assuming the maximum phonon number of the reconstructed state is $n_{\mathrm{max}}$, 
% we displace the $d$ modes along $d$ circles respectively,
we displace each mode along a circle,
\begin{align}
        \alpha_{j, p_j} = \left|\alpha_j\right|\exp\left[i\left(\pi/N\right)p_j\right],~j=1,\cdots,d
    \label{dispSet}
\end{align}
where $N=n_{\mathrm{max}}+1$ and $p_j\in\left\{-N,\cdots,N-1\right\}$. Then we perform a $d$-dimensional discrete Fourier transform of these $(2N)^d$ sets of $Q_{k_1,\cdots, k_d}\left(\alpha_{1,p_1},\cdots, \alpha_{d,p_d}\right)$ and truncate the summation at $n_{\textrm{max}}$:

\begin{align}
    Q_{k_1\cdots k_d}^{(l_1\cdots l_d)} 
    =& \frac{1}{(2N)^d}\sum_{p_1=-N}^{N-1}\cdots\sum_{p_d=-N}^{N-1}\nonumber\\
    % \sum_{q=-N}^{N-1}
    &\left[Q_{k_1\cdots k_d}(\alpha_{1,p_1},\cdots,\alpha_{d,p_d})
    e^{-i\sum_{j=1}^{d}\left(l_j p_j\right)\pi/N}\right]\nonumber\\
    =& \sum_{n_1=\mathrm{max}(0, -l_1)}^{n_{\textrm{max}}}\cdots
    \sum_{n_d=\mathrm{max}(0, -l_d)}^{n_{\textrm{max}}}
    \gamma_{k_1n_1}^{(l_1)}\cdots\gamma_{k_dn_d}^{(l_d)}\nonumber\\
    &\cdot\rho_{l_1+n_1,\cdots, l_d+n_d, n_1,\cdots, n_d}
    \label{FT}
\end{align}
% \end{widetext}
with 
\begin{align}
    \gamma_{k_in_i}^{(l_i)} =& \frac{e^{-\left|\alpha_i\right|^2}\left|\alpha_i\right|^{2k_i}}{k_i!}
    \sum_{j_{i}=0}^{\mathrm{min}\left(k_i, n_i+l_i\right)}
    \sum_{j_{i}'=0}^{\mathrm{min}\left(k_i, n_i\right)}
    \left|\alpha_i\right|^{2(n_i-j_i-j_i')+l_i}\nonumber\\
    &  \cdot(-1)^{-j_{i}-j_{i}'}{k_i\choose j_i}{k_i'\choose j_i'}\frac{\sqrt{n_i!(n_i+l_i)!}}{{(n_i-j_i')!(n_i+l_i-j_i)!}}
    \label{gamma}
\end{align}
where $i=1,\cdots,d$ and 
$\rho_{l_1+n_1,\cdots, l_d+n_d, n_1,\cdots, n_d}=\left<l_1+n_1, \cdots, l_d+n_d\left|\rho\right|n_1\cdots n_d\right>$ (See derivations in supplementary material). In Eqs.~(\ref{FT}), (\ref{gamma}) the values of $Q_{k_1,\cdots, k_d}\left(\alpha_{1,p_1},\cdots, \alpha_{d,p_d}\right)$ come from $d$-mode BSB fitting and $\gamma_{k_in_i}^{(l_i)}$ are known once the displacement distances $\left|\alpha_i\right|$ are fixed.
Therefore each element in the target state density matrix $\rho_{(l_1+n_1),\cdots, (l_d+n_d), n_1,\cdots, n_d}$ can be reconstructed \cite{Zhang2021thesis}. 
In reality we set a maximum cutoff $k_{\mathrm{max}}$ in Eq.~(\ref{dDBSB}) based on the expected input state maximum phonon number to avoid infinite summation, which satisfies $k_{\mathrm{max}}\geq n_{\mathrm{max}}$ since displacement operations can potentially increase the maximum phonon number of interest. 

\begin{figure}[htbp]
    \centering
    \includegraphics[width=\linewidth]{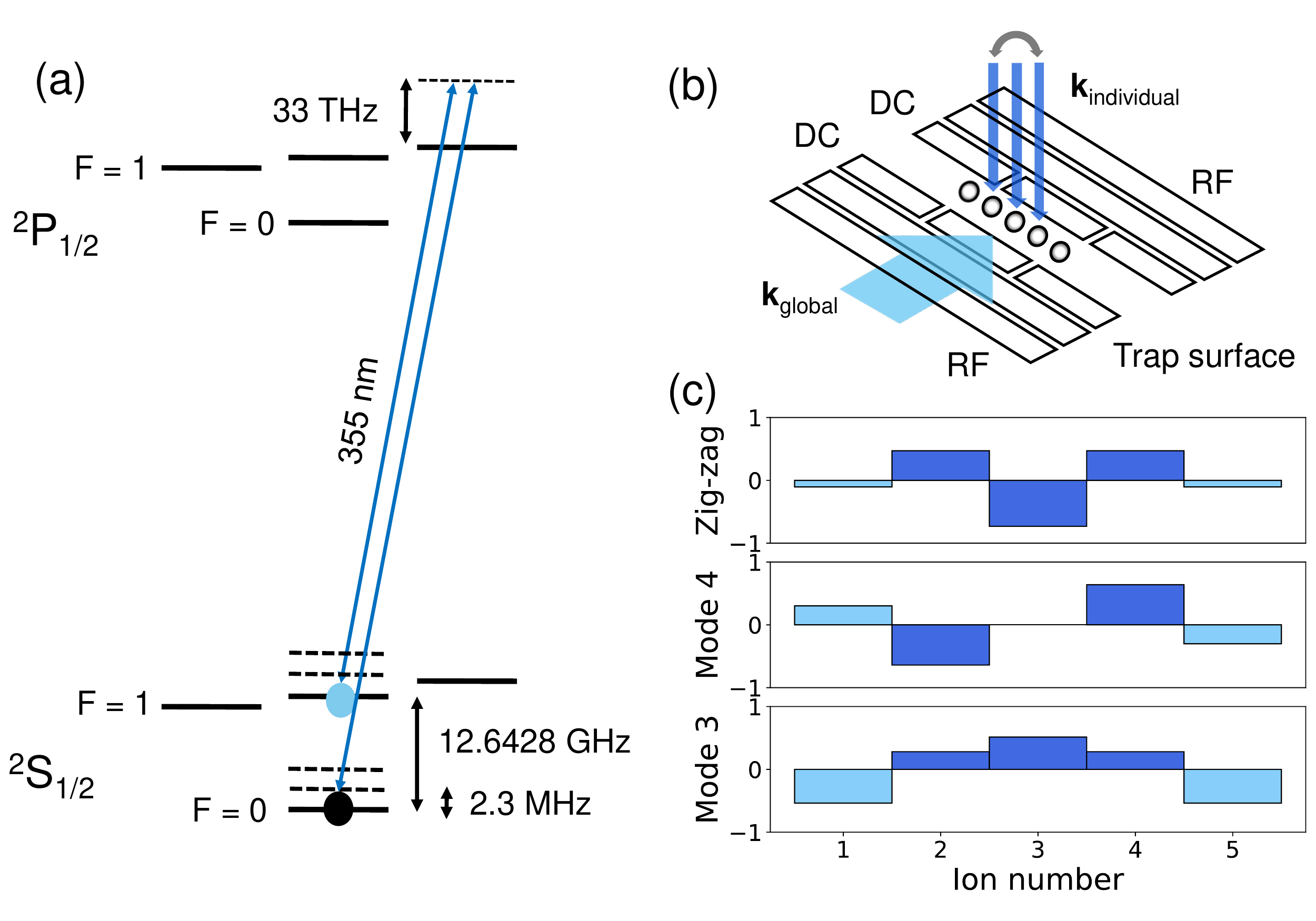}
    \caption{(a) Qubit and motional energy levels of $^{171}$Yb$^+$ ion. Two-photon Raman transitions of wavelength 355 nm are applied for qubit rotations and ion-motion coupling operations. (b) Schematic diagram of our surface trap and Raman beam configurations (not to scale). The two individual addressing beams are focused on the second and third in the chain. By tilting the MEMS mirrors the individual Raman beams can point to different ions. (c) Coupling strength of zig-zag mode, the fourth and the third radial mode to each ion, with the second, third and fourth ion highlighted in dark blue color.}
    \label{config}
\end{figure}

{\textit{Experiment.---}}We experimentally verify the $d$-mode BSB fitting method for $d=2$ and 3 and the density matrix reconstruction method for $d=2$. The experiment is conducted in a room temperature trapped-ion system~\cite{Wang2020}. As is shown in Fig.~\ref{config}, a 5-ion $^{171}$Yb$^{+}$ chain with an average ion separation of about 5 $\mu$m is confined in a micro-fabricated linear radio-frequency Paul trap~\cite{Revelle2020} with radial motional modes $\omega\sim2.3$ MHz. To avoid cross coupling between the modes and to reduce the heating effect, the zig-zag mode and the third radial mode are selected for 2-mode motional state preparation and measurement, with a frequency separation of about 85 kHz. 
For 3-mode motional states we also drive the fourth radial mode, which is about 50 kHz away from zig-zag mode.
The qubit levels are encoded in the hyperfine ground states $\left|\downarrow\right>\equiv~^2$S$_{1/2}\left|F=0, m_F=0\right>$ and $\left|\uparrow\right>\equiv~^2$S$_{1/2}\left|F=1, m_F=0\right>$ with a frequency splitting of $12.642821$ GHz~\cite{Olmschenk2007}. The qubit rotations and ion-motion coupling transitions are driven by stimulated Raman transitions using two orthogonal mode-locked 355 nm picosecond-pulsed laser beams. One elliptical global beam is shined on all ions and two individual addressing beams are tightly focused onto the second and third ions and can be steered across the chain to address the other ions using MEMS tilting mirrors. The amplitude, frequency and phase of Raman beams are controlled by a Xilinx Zync UltraScale+ Radio Frequency System-on-Chip board using firmware designed by Sandia National Laboratories~\cite{RFSoC}. A multi-channel fiber array connected to separated photomultiplier tubes is used to detect fluorescence from each ion, enabling individual detection and joint spin state distribution measurement of multiple ions. 

\begin{figure*}[htbp]
    \centering
    \includegraphics[width=\linewidth]{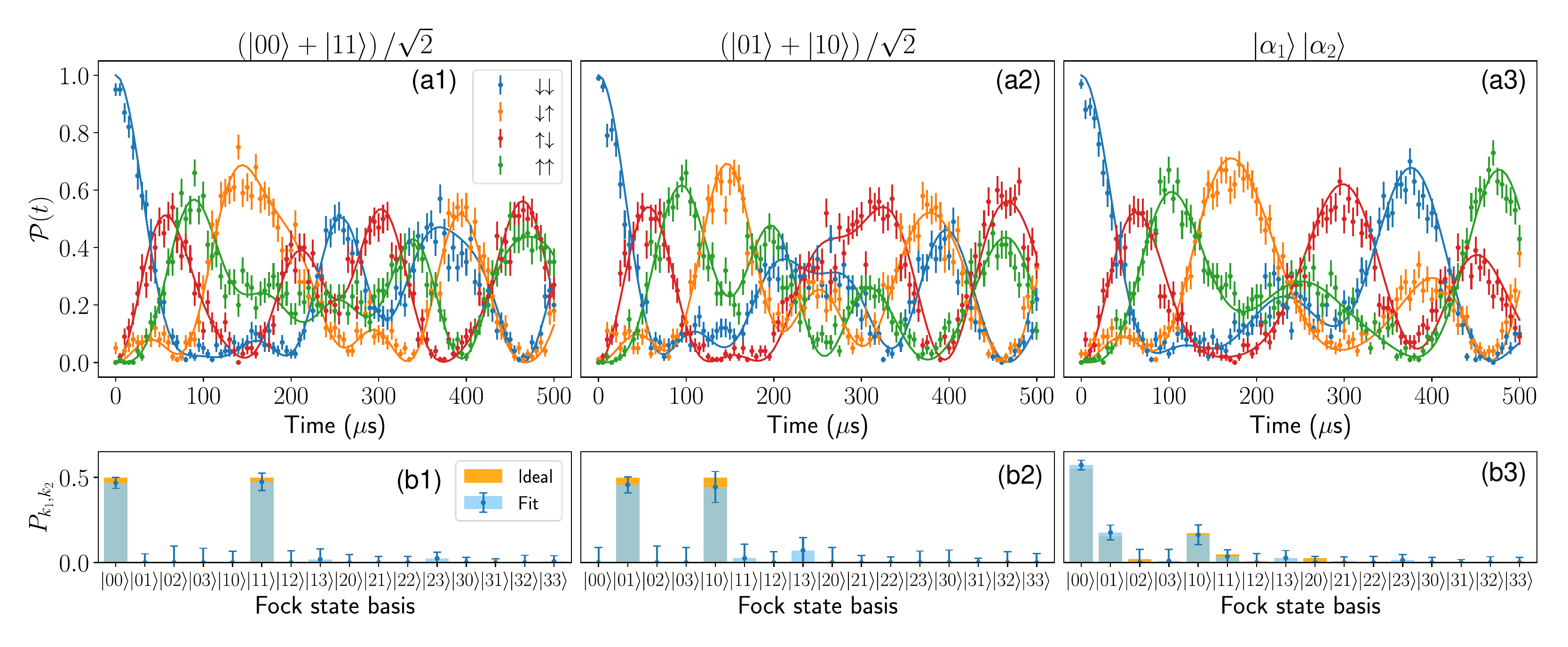}
    \caption{2-mode BSB fitting results for $\left(\left|00\right>+\left|11\right>\right)/\sqrt{2}$, $\left(\left|01\right>+\left|10\right>\right)/\sqrt{2}$, and  $\left|\alpha_1\right>\left|\alpha_2\right>$. In (a1) (a2) and (a3), the joint spin state distributions $\mathcal{P}_{\downarrow\downarrow}$, $\mathcal{P}_{\downarrow\uparrow}$, $\mathcal{P}_{\uparrow\downarrow}$, $\mathcal{P}_{\uparrow\uparrow}$ (dots) and the curve of ideal case (lines) are plotted. Each point is averaged over 100 experiments and the errorbars denote one standard deviation. The Fock state distribution fitting results with uncertainties (blue) are shown in (b1) (b2) and (b3) along with the ideal case (orange). Within the uncertainties, the fitting values of $P_{k_1, k_2}$ match with the expected values.}
    \label{diag}
\end{figure*}

At the start of the experiment, the ions are laser cooled using Doppler and EIT cooling to $\bar{n}\approx 0.3$, then the modes used for experiment are further sideband cooled to the motional ground state {($\bar{n}\approx0.03$)}. The 2-mode target states are prepared by driving carrier, motional sideband transitions and push operations on the third ion (See supplementary material). The 3-mode target state is prepared using the second ion since all three motional modes are coupled to the second ion.
% the fourth radial mode is decoupled to the third ion.
The coherent displacement operation with controlled distance and phase is realized by applying the Hamiltonian in Eq.~(\ref{diag}) with the ion spin rotated to the +1-eigenstate of Eq.~(\ref{diag}).
The displacement distance is calibrated by applying the displacement operator to a motional ground state, then measuring the Cummings collapse and revival of BSB transition and fitting to a coherent state~\cite{Meekhof1996}. After applying the coherent displacement operation the ion spin is rotated back to $\left|\downarrow\right>$ state. 
Finally we drive the 2(3)-mode BSB transition and all joint spin state distributions are recorded. 
The 2-mode BSB transition is driven by individual Raman beams addressing the second and third ions, and the 3-mode BSB transition requires steering one individual Raman beam onto the fourth ion. Once the sideband Rabi frequencies of different modes on different ions are calibrated, the Fock state distributions $P_{k_1,\cdots, k_d}$ of either the target state or the displaced target state can be found using the $d$-mode BSB fitting method as described. 

We demonstrate the 2-mode BSB fitting by measuring the Fock state distribution of various 2-mode motional states. In Fig.~\ref{diag} we show the 2-mode BSB time scan curves and Fock state distribution fitting results of motional Bell states $\left(\left|00\right>+\left|11\right>\right)/\sqrt{2}$, $\left(\left|01\right>+\left|10\right>\right)/\sqrt{2}$, and the product of two coherent states $\left|\alpha_1\right>\left|\alpha_2\right>$, where $\left|\alpha_1\right|= 0.56$ and $\left|\alpha_2\right|= 0.53$ are fitted from single-mode BSB time scan curve. Here we truncate at $k_{\mathrm{max}}=3$ to cover most non-zero Fock state distributions while not overfitting the data. The uncertainties of $P_{k_1, k_2}$ are extracted from the covariance matrix of least square fitting. The fitting result of $P_{k_1, k_2}$ shows an agreement with the expected values within one standard deviation, thereby proving that the 2-mode BSB fit method works properly.

\begin{figure}[htbp]
    \centering
    \includegraphics[width=\linewidth]{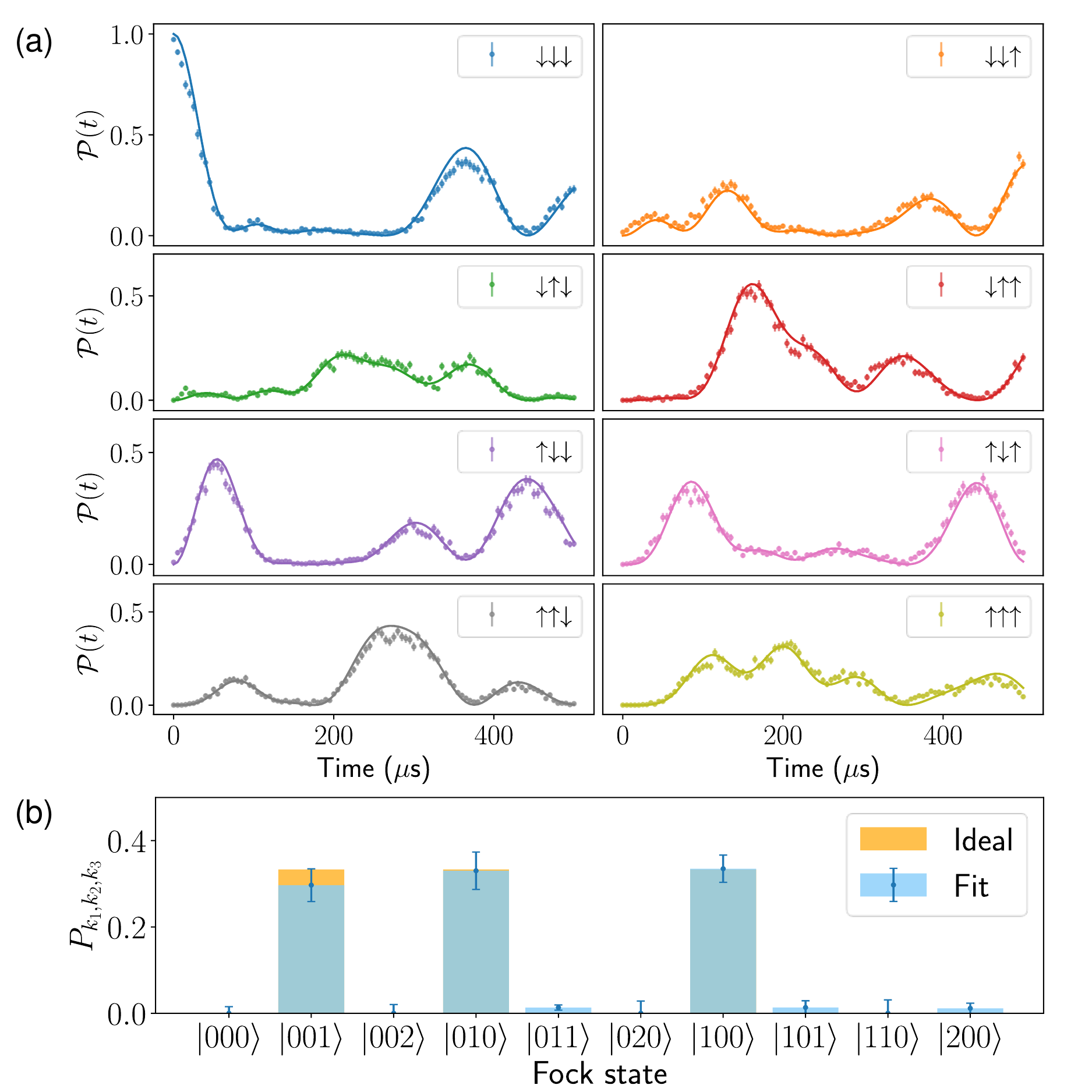}
    \caption{3-mode BSB time scan and fitting results for a motional W-state $\left(\left|100\right>+\left|010\right>+\left|001\right>\right)/\sqrt{3}$. In (a), the time scan curve for all 8 configurations of the three-ion joint spin state distributions are plotted together with the ideal curve and each data point is averaged over 400 experiments. The Fock state population fitting results along with infidelities (blue) and the ideal case (orange) are show in (b). The fitting results show a population close to $1/3$ in $\left|001\right>$, $\left|010\right>$ and $\left|100\right>$, and all other state populations are close to 0.}
    \label{WState}
\end{figure}

To verify the 3-mode BSB fitting, we prepare a motional W-state $\left(\left|100\right>+\left|010\right>+\left|001\right>\right)/\sqrt{3}$ then measure the 3-mode Fock state probabilities. Fig.~(\ref{WState}) shows the time scan curve and the Fock state distribution fitting results. For the fitting we select the Fock state basis that is at most one phonon different from $\left|100\right>$, $\left|010\right>$ and $\left|001\right>$ to avoid overfitting, and the fitting result shows an agreement with the ideal case within one standard deviation. In the Supplementary Materials, we use parity scans to show the entangled nature of the state. This result verifies that for higher number of motional modes the multi-mode BSB fitting can still extract the Fock state distributions. To reduce the uncertainties of the Fock state distribution result, one can repeat the experimental sequence for more times to obtain data sets with smaller errorbar and less influence from random fluctuations.

\begin{figure}[htbp]
    \centering
    \includegraphics[width=\linewidth]{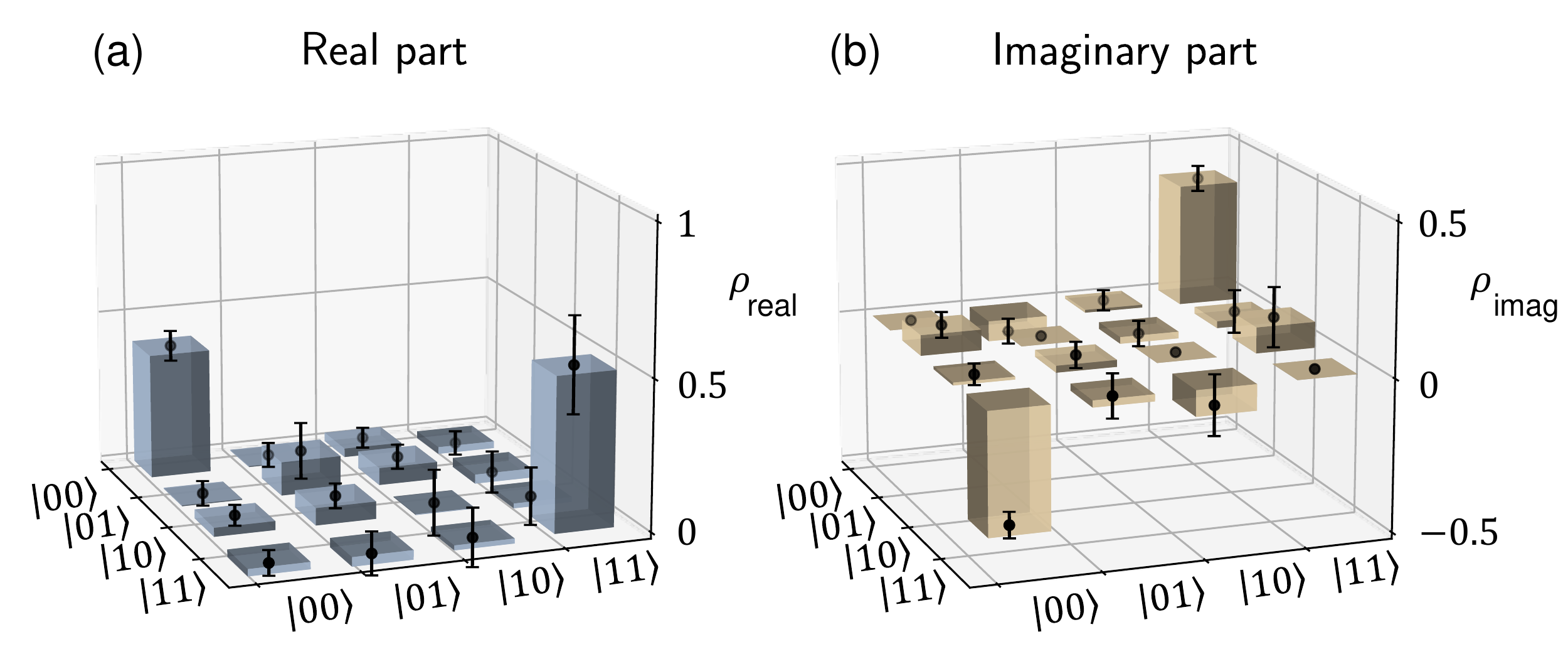}
    \caption{Reconstructed density matrix of $(\left|00\right>+i\left|11\right>)/\sqrt{2}$. The state was displaced by $\left|\alpha_1\right|=0.52$ and $\left|\alpha_2\right|=0.51$. In the real part (a) we observe a population of close to 0.5 on $\rho_{00, 00}$ and $\rho_{11, 11}$ while in the imaginary part (b) the off-diagonal term $\rho_{00, 11}$ and $\rho_{11, 00}$ has opposite signs and amplitudes close to 0.5. All other components in real and imaginary part are close to 0.}
    \label{DM}
\end{figure}

We further reconstruct the density matrix of motional Bell state $\left(\left|00\right>+i\left|11\right>\right)/\sqrt{2}$ up to $n_{\mathrm{max}}=1$. To realize this measurement we apply 16 pairs of displacement operations on 2 modes with different angles in phase spaces then measure phonon state populations $Q_{k_1, k_2}\left(\alpha_1, \alpha_2\right)$. From all values of $Q_{k_1, k_2}\left(\alpha_1, \alpha_2\right)$, we use Eq.~(\ref{FT})(\ref{gamma}) to reconstruct the density matrix. In Fig.~\ref{DM} we show the real and imaginary part of the reconstructed density matrix of $\left(\left|00\right>+i\left|11\right>\right)/\sqrt{2}$ with the uncertainties of each element. The reconstructed density matrix has the expected behavior $\rho_{00, 11}=-\rho_{11, 00}= 0.40(4) i$, $\rho_{00,00}=0.41(5)$ and $\rho_{11, 11}= 0.51(12)$, and achieves a fidelity of 0.87(11) including all state preparation and measurement error. The result shows a significant entanglement between two motional modes, which cannot be extracted from projection measurements on a single mode. Notice that the reconstruction process does not guarantee the positive semi-definiteness of final density matrix, which is mostly caused by non-ideal fitting of $Q_{k_1, k_2}\left(\alpha_1, \alpha_2\right)$. If we calculate the covariance matrix of the final density matrix elements based on that of 2-mode BSB fitting and extract the uncertainties, then within the uncertainties we can find a positive semi-definite density matrix that is closest to the reconstructed state.

{\textit{Outlook.---}}Generally, to extract the Fock state populations, the number of fitting parameters scales as $O\left((k_{\mathrm{max}})^d\right)$, and to reconstruct the full density matrix, $\left(2n_{\mathrm{max}}+2\right)^d$ sets of displacement operations with different push directions in phase spaces are required. In practice, as $k_\mathrm{max}$, $n_\mathrm{max}$ and $d$ scale up, the risk of overfitting increases, thus introducing larger uncertainties in the final reconstructed density matrix. However, the total number of parameters for reconstruction has the same order of magnitude as that of free parameters in a $d$-mode motional density matrix with maximum phonon number cutoff $n_{\mathrm{max}}$, $\left(n_{\mathrm{max}}\right)^{2d}-1$, therefore this reconstruction method is efficient. 

The reconstruction method can be implemented in any system that has Jaynes-Cummings type interactions, such as circuit QED~\cite{Wang2011, su2014fast} and optomechanical systems~\cite{Aspelmeyer2014, wollack2021quantum}. For the experiments in which multi-mode Fock state distribution is of interest, such as phononic boson sampling~\cite{Shen2014}, the multi-mode BSB fitting method can be of great benefit.

\begin{acknowledgements}
We thank Shilin Huang and Brad Bondurant for helpful discussion. This work is supported by the Office of the Director of National Intelligence - Intelligence Advanced Research Projects Activity through ARO Contract No. W911NF-16-1-0082 (experimental setup), National Science Foundation STAQ Project No. Phy-181891 (trapped-ion simulation), the U.S. Department of Energy (DOE), Office of Advanced Scientific Computing Research award DE-SC0019294 (trapped-ion tomography), DOE Basic Energy Sciences Award DE-0019449 (spin-motion pulses) and ARO MURI grant W911NF-18-1-0218 (distributed measurement protocol).
\end{acknowledgements}

\bibliographystyle{apsrev4-1} % Tell bibtex which bibliography style to use
\bibliography{./mybib}

% \newpage
\clearpage

\onecolumngrid
\section{Supplementary Materials}
\subsection{Target State Preparation and Phase Control}

The various motional target states are prepared by implementing precise control over frequency, amplitude, phase and pulse length of Raman laser beams~\cite{Meekhof1996, Leibfried1996}. In experiment the control is realized by using the RFSoC to generate RF signals with certain frequency, amplitude and phases, then sending sequences of signals to acoustic-optical modulators (AOMs) to control Raman lasers then realize different operations on ions~\cite{RFSoC}. A list of operations that are used in motion state preparations and measurements are denoted as follows:
\begin{itemize}
    \item $R_C\left(\theta, \phi\right)$: Carrier rotation of angle $\theta$ and phase $\phi$
    \item $R_{BSB, j}\left(\theta, \phi\right)$: Blue-sideband rotation on mode $j$ of angle $\theta$ and phase $\phi$
    \item $D_{+_i, j}\left(\alpha\right)$: Spin-dependent push on mode $j$, with spin eigenstate $\left|+i\right>=\left(\left|\downarrow\right>+i\left|\uparrow\right>\right)/\sqrt{2}$ and motion displacement amount $\left|\alpha\right|$ and angle $\mathrm{arg}\left(\alpha\right)$ in phase space
\end{itemize}

Here all phases $\phi$ are referenced to how we experimentally define the phase of spin-$\left|+i\right>$ state, and spin-dependent push is implemented by on-resonance driving red- and blue-sideband transitions simultaneously with phases $\phi_{r,b}$ and the same sideband Rabi frequencies $\Omega$ as in Eq.~(\ref{disp}).
% \begin{align}
%     H &= -i\Omega\left(\sigma_+a e^{i\phi_r}-\sigma_-a^\dagger e^{-i\phi_r}+\sigma_+a^\dagger e^{i\phi_b}-\sigma_-a e^{-i\phi_b}\right)\nonumber\\
%     &= -i\Omega\left(\sigma_+e^{i\phi_s}-\sigma_-e^{-i\phi_s}\right)\left(a^\dagger e^{i\phi_m}+ae^{-i\phi_m}\right)
% \end{align}
where $\phi_{s} = \left(\phi_b+\phi_r\right)/2$ and $\phi_{m} = \left(\phi_b-\phi_r\right)/2$. In order to realize coherent displacement operations on motion states we need to calibrate $\phi_{s}$ such that the eigenstate of $-i\left(\sigma_+e^{i\phi_s}-\sigma_-e^{-i\phi_s}\right)$ is $\left|+i\right>$ state, which theoretically should be $0^{\circ}$.

\begin{figure}[htbp]
    \centering
    \includegraphics[width=0.5\linewidth]{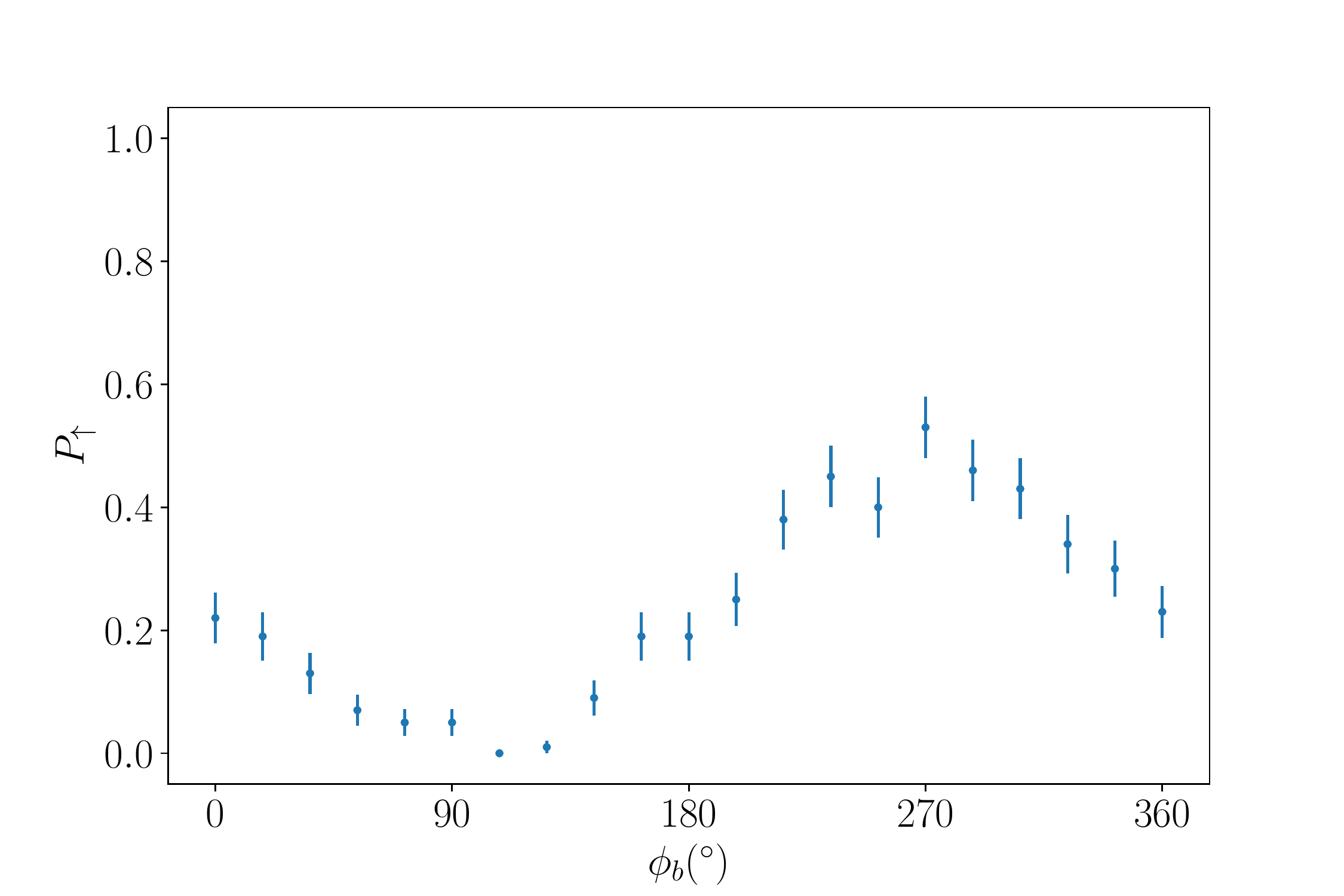}
    \caption{Phase calibration between single-tone rotation and two-tone push, which is achieved by $R_C\left(\pi/2, 0\right)\rightarrow $state-dependent push$\rightarrow R_C\left(-\pi/2, 0\right)$. We observe that at $\phi_r=0^\circ$ and $\phi_b=110^\circ$ the spin is rotated back to $\left|\downarrow\right>$ state and it holds for both motional modes.}
    \label{calibration}
\end{figure}

To calibrate $\phi_s$, we first prepare the spin to $\left|+i\right>$ state by applying $R_{C}({\pi}/{2}, 0)$, then apply a spin-dependent push operation with $\phi_{r}=0$ and varying $\phi_b$, and apply $R_{C}(-{\pi}/{2}, 0)$ then measure the spin. If $\left|+i\right>$ is the eigenstate of $\left(\sigma_+e^{i\phi_s}-\sigma_-e^{-i\phi_s}\right)$, then the final state is $\left|\downarrow\right>\left|\alpha\right>$ and we should see the spin rotated back to $\left|\downarrow\right>$, otherwise we create a spin-motion cat state and measurement on spin will go up eventually to 0.5. As is shown in Fig.~(\ref{calibration}), we observe that at $\phi_{b}=110^{\circ}$ the spin goes down to $\left|\downarrow\right>$ state, indicating $\phi_s$ is synchronized to the phase of single-qubit rotation $R_{C}({\pi}/{2}, 0)$ at $\phi_s=55^\circ$. The mismatch with theoretical value comes from a certain phase offset in our control hardware between single-tone and two-tone output.

We then can experimentally prepare 2-mode motional Bell states, products of coherent states and 3-mode motional W-state using the following pulse sequences (global phase factors are ignored):
% (phases of $\pi$-rotations do not affect target states and are thus ignored):

\begin{align}
    \left|\downarrow\right>\left|0_1\right>\left|0_2\right>&\xrightarrow{R_{BSB1}(\frac{\pi}{2}, \phi_1)}\frac{1}{\sqrt{2}}\left(\left|\downarrow\right>\left|0_1\right>\left|0_2\right>+e^{i\phi_1}\left|\uparrow\right>\left|1_1\right>\left|0_2\right>\right)\xrightarrow{R_{RSB2}(\pi,\phi_2)}\frac{1}{\sqrt{2}}\left|\downarrow\right>\left(\left|0_1\right>\left|0_2\right>+e^{i(\phi_1+\phi_2)}\left|1_1\right>\left|1_2\right>\right)\\
    \left|\downarrow\right>\left|0_1\right>\left|0_2\right>&\xrightarrow{R_{BSB1}(\frac{\pi}{2}, \phi_1)}\frac{1}{\sqrt{2}}\left(\left|\downarrow\right>\left|0_1\right>\left|0_2\right>+e^{i\phi_1}\left|\uparrow\right>\left|1_1\right>\left|0_2\right>\right)\xrightarrow{R_{BSB2}(\pi,\phi_2)}\frac{1}{\sqrt{2}}\left|\uparrow\right>\left(e^{i\phi_2}\left|0_1\right>\left|1_2\right>+e^{i\phi_1}\left|1_1\right>\left|0_2\right>\right)\nonumber\\
    &\xrightarrow{R_{C}(\pi)}\frac{1}{\sqrt{2}}\left|\downarrow\right>\left(e^{i\phi_2}\left|0_1\right>\left|1_2\right>+e^{i\phi_1}\left|1_1\right>\left|0_2\right>\right)\\
    \left|\downarrow\right>\left|0_1\right>\left|0_2\right>&\xrightarrow{R_{C}(\frac{\pi}{2}, 0)}\left|+i\right>\left|0_1\right>\left|0_2\right>\xrightarrow{D_{+_i,1}\left(\alpha_1\right)}\left|+i\right>\left|\alpha_1\right>\left|0_2\right>\xrightarrow{D_{+_i,2}\left(\alpha_2\right)}\left|+i\right>\left|\alpha_1\right>\left|\alpha_2\right>\xrightarrow{R_{C}(-\frac{\pi}{2}, 0)}\left|\downarrow\right>\left|\alpha_1\right>\left|\alpha_2\right>\\
    \left|\downarrow\right>\left|0_1\right>\left|0_2\right>\left|0_3\right>&\xrightarrow{R_{BSB1}(\frac{\pi}{3}, \phi_1)}\sqrt{\frac{2}{3}}\left|\downarrow\right>\left|0_1\right>\left|0_2\right>\left|0_3\right>+\frac{1}{\sqrt{3}}e^{i\phi_1}\left|\uparrow\right>\left|1_1\right>\left|0_2\right>\left|0_3\right>\nonumber\\
    &\xrightarrow{R_{BSB2}(\frac{\pi}{2}, \phi_2)}\frac{1}{\sqrt{3}}\left|\downarrow\right>\left|0_1\right>\left|0_2\right>\left|0_3\right>+\frac{1}{\sqrt{3}}e^{i\phi_1}\left|\uparrow\right>\left|1_1\right>\left|0_2\right>\left|0_3\right>+\frac{1}{\sqrt{3}}e^{i\phi_2}\left|\uparrow\right>\left|0_1\right>\left|1_2\right>\left|0_3\right>\nonumber\\
    &\xrightarrow{R_{BSB3}(\pi,\phi_3)}\frac{1}{\sqrt{3}}\left|\uparrow\right>\left(e^{i\phi_3}\left|0_1\right>\left|0_2\right>\left|1_3\right>+e^{i\phi_1}\left|1_1\right>\left|0_2\right>\left|0_3\right>+e^{i\phi_2}\left|0_1\right>\left|1_2\right>\left|0_3\right>\right)\nonumber\\
    &\xrightarrow{R_{C}(\pi)}\frac{1}{\sqrt{3}}\left|\downarrow\right>\left(e^{i\phi_3}\left|0_1\right>\left|0_2\right>\left|1_3\right>+e^{i\phi_1}\left|1_1\right>\left|0_2\right>\left|0_3\right>+e^{i\phi_2}\left|0_1\right>\left|1_2\right>\left|0_3\right>\right)
    \label{Wprep}
\end{align}

After target state preparation, we prepare the spin to $\left|+_i\right>$ state then apply the set of displacements in Eq.~(\ref{dispSet}) by adding and subtracting the same phase on the calibrated $\phi_b$ and $\phi_r$, which changes $\phi_m$ without affecting $\phi_s$. Notice that when reconstructing the target state $\left(\left|00\right>+e^{i\varphi}\left|11\right>\right)/\sqrt{2}$, the phase $\varphi$ is defined differently: $\varphi$ is instead the relative phase between the prepared Bell state and the default motional phase of the set of displacement operations, $\phi_m$. Our calibration regime introduces a default $\phi_m=55^\circ$ which is taken into account when reconstructing the target state.

% \nbar for different modes after sideband cooling
\subsection{Sideband Cooling for Motional Modes}

\begin{figure}[htbp]
    \centering
    \includegraphics[width=0.8\textwidth]{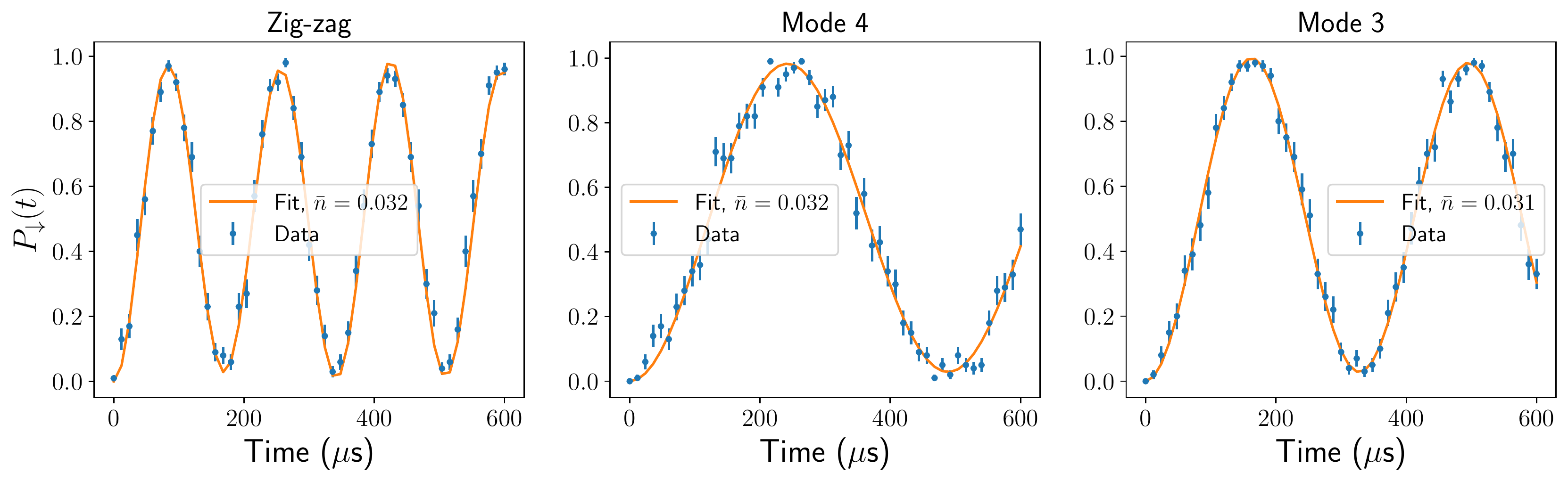}
    \caption{BSB time scan of all 3 modes after sideband cooling. All modes are cooled to $\bar{n}\approx 0.03$.}
    \label{SBCool}
\end{figure}

The third, fourth and zig-zag mode in our 5-ion chain has negligible heating rate~\cite{Wang2020}. At the start of our experiment, we apply Doppler and EIT cooling to all motional modes, then apply sideband cooling to cool the modes we want to motional ground state. To verify the cooling effect, we drive BSB time scans on each mode after sideband cooling, then fit the BSB time scan curve with a thermal distribution. As is shown in Fig.~(\ref{SBCool}), all three modes after sideband cooling are cooled to $\bar{n}\approx 0.03$.

\subsection{Error for Reconstruction}

\begin{figure}[htbp]
    \centering
    \includegraphics[width=\textwidth]{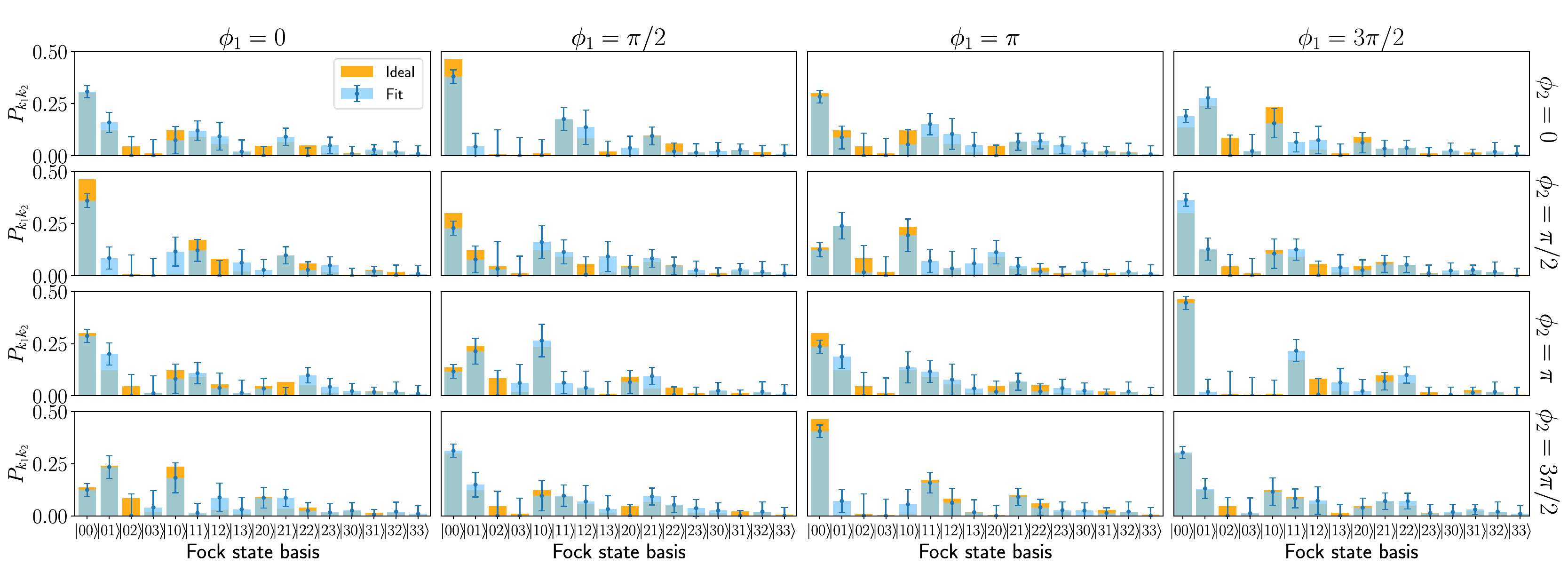}
    \caption{The Fock state distribution fitting results for displaced target state $\left(\left|00\right>+i\left|11\right>\right)/\sqrt{2}$. To reconstruct the target state up to $n_{\mathrm{max}}=1$, we displace two modes following angles in Eq.~(\ref{dispSet}), generating 16 sets of $Q_{k_1, k_2}\left(\alpha_1, \alpha_2\right)$. Most fit values agree with the ideal values within one standard deviation.}
    \label{PushedDiag}
\end{figure}

% We blame drifting.
To reconstruct the target motional density matrix, we fit the $d$-mode BSB time scan to extract the Fock state distributions of displaced target state $Q_{k_1\cdots k_d}\left(\alpha_1,\cdots, \alpha_d\right)$, then apply $d$-dimensional Fourier transform to reconstruct the density matrix components. Fig.~(\ref{PushedDiag}) shows our fit result of $Q_{k_1, k_2}\left(\alpha_1, \alpha_2\right)$ when reconstructing $\left(\left|00\right>+i\left|11\right>\right)/\sqrt{2}$. We observe a difference between fit result and ideal case, which contributes to the non-ideal reconstructed state fidelity. However, most of the fitting results match to the ideal case within one standard deviation. We expect that with a more precise calibration we can achieve higher reconstructed state fidelity.

After reconstructing the density matrix as is shown in Fig.~(\ref{DM}), we search for a closest positive semi-definite density matrix to check whether it is within the errorbar of the result. The reconstructed density matrix with uncertainties is
\begin{align}
\rho =
    \begin{bmatrix}
    0.41(5) & 0.00(4)-0.07(4)i & 0.03(3)+0.00(3)i & -0.02(4)+0.40(4)i\\
    0.00(4)+0.07(4)i & 0.11(9) & 0.05(4)-0.02(4)i & -0.03(7)+0.02(7)i\\
    0.03(3)-0.00(3)i & 0.05(4)+0.02(4)i & 0.0(1)  & -0.02(9)+0.08(9)i\\
    -0.02(4)-0.40(4)i& -0.03(7)-0.02(7)i & -0.02(9)-0.08(9)i & 0.51(12)
    \end{bmatrix}
\end{align}
and the closest positive semi-definite density matrix is
\begin{align}
\rho_{\mathrm{ps}} =
    \begin{bmatrix}
    0.38 & 0-0.06i & 0.03 & -0.02+0.37i\\
    0+0.06i & 0.11 & 0.04-0.02i & -0.03+0.02i\\
    0.03 & 0.04+0.02i & 0.03  & -0.01+0.07i\\
    -0.02-0.37i& -0.03-0.02i & -0.01-0.07i & 0.48
    \end{bmatrix}
\end{align}
which has a trace distance to the reconstructed density matrix of 0.06 and is within the errorbar of the reconstructed density matrix.

% W-state witness
\subsection{Verifying W-state Entanglement}

From the 3-mode BSB time scan measurement in Fig.~(\ref{dDBSB}) we can extract the diagonal terms in the prepared W-state density matrix, however the measurement does not inform knowledge of off-diagonal terms which describes the entanglement of W-state. Reconstructing the whole density matrix of W-state can extract the entanglement, which requires long measurement process and can be affected by show drift in calibrations. Here we propose another method to verify the W-state entanglement. 

After preparing W-state using pulse sequences in Eq.~(\ref{Wprep}), we apply a carrier $\pi$-pulse to flip the ion spin to $\left|\uparrow\right>$, then apply $R_{BSB,i}\left(\pi, \phi_1\right)$ followed by a $R_{BSB,j}\left(\pi/2, \phi_2\right)$, with $i, j\in\{1, 2, 3\}$ and $i\neq j$. Here we assume all motional Fock states except $\left|001\right>$, $\left|010\right>$ and $\left|100\right>$ are unoccupied, which holds according to the Fock state distribution measurement.

\begin{figure}[htbp]
    \centering
    \includegraphics[width=0.8\textwidth]{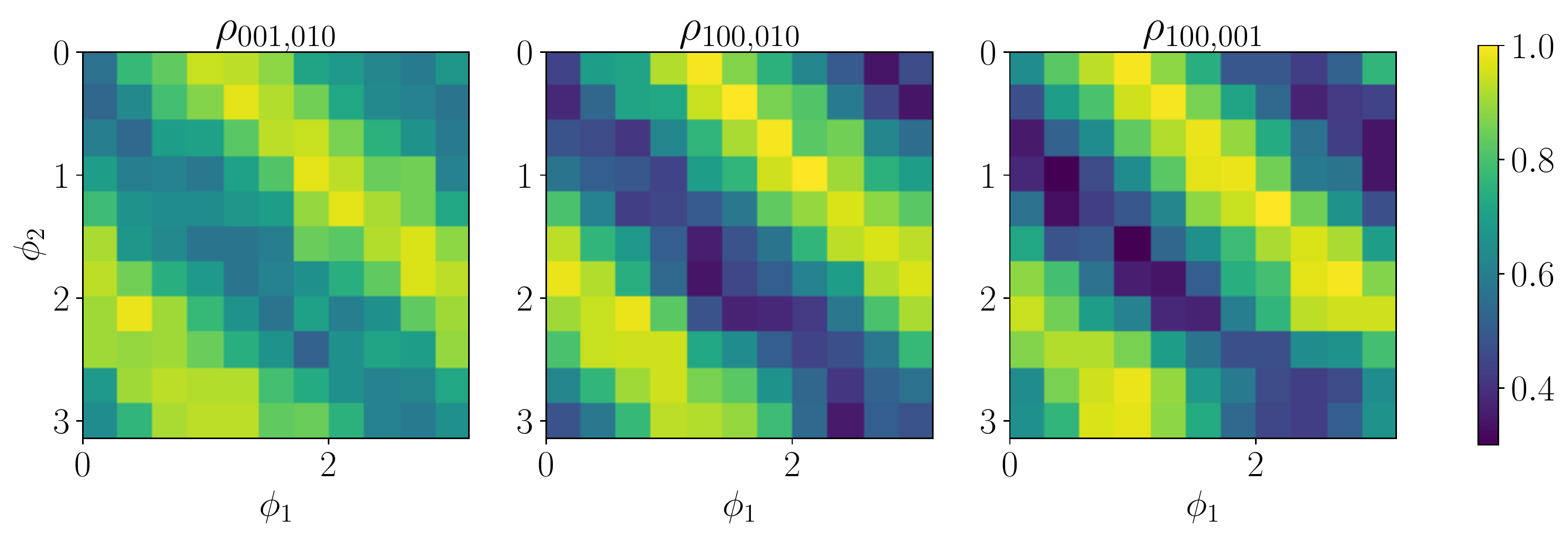}
    \caption{Phase scan patterns of all 3 pairs of modes. The pattern verifies the entanglement of prepared W-state.}
    \label{parity}
\end{figure}

For simplicity we define:
\begin{align}
    \textrm{Fock(1)} = 001,~\textrm{Fock(2)} = 010,~\textrm{Fock(1)} = 100
\end{align}
The probability of observing the ion in $\left|\downarrow\right>$ after applying the pulse sequence is
\begin{align}
    \mathcal{P}_{ij}(\downarrow) = P_{\textrm{Fock}(i)}+P_{\textrm{Fock}(j)}+\left|\left<\mathrm{Fock}(i)\left|\rho\right|\mathrm{Fock}(j)\right>\right|\cos(\phi_1+\phi_2)
\end{align}
Therefore by scanning $\phi_1$ and $\phi_2$ we can verify the entanglement with the sinusoidal pattern. Fig.~(\ref{parity}) shows the pattern of all 3 pairs of phase scans thus verifies the W-state entanglement.

\subsection{$d$-mode Discrete Fourier Transform}

Here we extended the result in~\cite{Leibfried1996} to $d$-mode case provided that the Fock state distributions can be measured. Relevant discussion can be found in Appendix A of Ref. \cite{Zhang2021thesis}. We present it here for convenience.

We define the Fock state distributions of displaced target state as
\begin{align}
    Q_{k_1\cdots k_d} (\alpha_1,\cdots, \alpha_d)
     & = \left<k_1\cdots k_d\left|\prod_{i=1}^d D_i^{\dag}(\alpha_i) \rho \prod_{i=1}^d D_i(\alpha_i)\right|k_1\cdots k_d\right>\nonumber\\
    & = \left<0\cdots 0\left|\left(\prod_{i=1}^d\frac{(a_i)^{k_i}D_i^{\dag} (\alpha_i)}{\sqrt{k_i!}} \right)\rho \left(\prod_{i=1}^d\frac{D_i(\alpha_i)(a_i^\dag)^{k_i} }{\sqrt{k_i!}} \right)\right|0\cdots 0\right>
\end{align}

Using the relation $D^{^{\dag}} (\alpha) a D (\alpha) = a + \alpha$ and definition of coherent state $\left|\alpha\right>=e^{-\frac{|\alpha|^2}{2}} \sum_{n=0}^{\infty}\frac{\alpha^n}{\sqrt{n!}}\left|n\right>$ we obtain
\begin{align}
    Q_{k_1\cdots k_d} (\alpha_1,\cdots, \alpha_d) =& 
    \left(\prod_{i=1}^d\frac{e^{-|\alpha_i|^2}}{k_i!}\right)
    % \frac{1}{k_1!\cdots k_d !} e^{- | \alpha_1 |^2 -\cdots- | \alpha_d |^2} 
    \sum_{n_1,\cdots, n_d,n'_1,\cdots, n'_d = 0}^{\infty}
    \left(\prod_{i=1}^d\frac{(\alpha_i^{*})^{n_i}}{\sqrt{n_i !}}\frac{(\alpha_i)^{n'_i}}{\sqrt{n'_i !}}\right)
    % \frac{(\alpha_1^{*})^{n_1}}{\sqrt{n_1 !}} \frac{(\alpha_2^{*})^{n_2}}{\sqrt{n_2 !}}\frac{(\alpha_1)^{n'_1}}{\sqrt{n'_1 !}} \frac{(\alpha_2)^{n'_2}}{\sqrt{n'_2 !}}
    \nonumber\\
    &~\cdot\left<n_1\cdots n_d\left| 
    \left(\prod_{i=1}^d(a_i - \alpha_i)^{k_i}\right) \rho
    \left(\prod_{i=1}^d(a_i^{\dag} - \alpha^{*})^{k_i}\right) \right|n'_1\cdots n'_d \right>
\end{align}

Then we apply binomial expansion to expand the equation above
\begin{align}
    Q_{k_1\cdots k_d} (\alpha_1,\cdots, \alpha_d) = &
    \left(\prod_{i=1}^d\frac{e^{-|\alpha_i|^2}|\alpha_i |^{2 k_i}}{k_i!}\right)
    \sum_{n_1,\cdots n_d, n'_1,\cdots, n'_d = 0}^{\infty} 
    \sum_{j_1,  j_1' = 0}^{k_1}\cdots \sum_{j_d, j_d' = 0}^{k_d}
    \rho_{n_1 + j_1,\cdots, n_d + j_d, n'_1 + j'_1,\cdots, n'_d + j_d'}
    \nonumber\\ 
  &~\cdot\left(\prod_{i=1}^d(-1)^{-j_i-j'_i}\frac{(\alpha_i^{*})^{n_i - j'_i}}{n_i !}\frac{(\alpha_i)^{n'_i -
  j_i}}{n'_i !}{k_i\choose j_i}{k_i\choose j'_i}\sqrt{ (n_i + j_i) ! (n'_i + j'_i) !}\right)
\end{align}
here $\rho_{n_1 + j_1,\cdots, n_d + j_d, n'_1 + j'_1,\cdots, n'_d + j_d'}=\left<n_1+j_1,\cdots,n_d+j_d\left|\rho\right|n'_1+j'_1,\cdots,n'_d+j'_d\right>$.

We apply $(2N)^d$ sets of displacement operations onto $d$ motional modes with displacement amount and angle as in Eq.~(\ref{dispSet}). After extracting all $Q_{k_1\cdots k_d} (\alpha_1,\cdots, \alpha_d)$ values we perform a $d$-mode discrete Fourier transform:
\begin{align}
    Q_{k_1\cdots k_d}^{(l_1\cdots l_d)}  = & \frac{1}{(2N)^d} \left(\prod_{i=1}^d\frac{e^{-|\alpha_i|^2}|\alpha_i |^{2 k_i}}{k_i!}\right)
  \sum_{n_1,\cdots n_d, n'_1,\cdots, n'_d = 0}^{\infty} 
    \sum_{j_1,  j_1' = 0}^{k_1}\cdots \sum_{j_d, j_d' = 0}^{k_d}\rho_{n_1 + j_1,\cdots, n_d + j_d, n'_1 + j'_1,\cdots, n'_d + j_d'}\nonumber\\
   & 
   \left(\prod_{i=1}^d(-1)^{-j_i-j'_i}\frac{(\alpha_i^{*})^{n_i - j'_i}}{n_i !}\frac{(\alpha_i)^{n'_i -
  j_i}}{n'_i !}{k_i\choose j_i}{k_i\choose j'_i}\sqrt{ (n_i + j_i) ! (n'_i + j'_i) !}\right)\nonumber\\
  &\cdot\sum_{p_1 = - N}^{N - 1}\cdots\sum_{p_d = - N}^{N - 1} \exp \left[ - i 
  \sum_{j=1}^d\left(p_j(l_j-m_j+m'_j)\pi / N\right)
%   \left(p (l_1 - m_1 + m_1')+q (l_2 - m_2  + m_2')\right)
  \right]
  \label{FT_-1}
\end{align}
assuming $n_i + j_i = m_i$ and $n_i' + j_i' = m_i'$ for $i=1,\cdots, d$. The value of $Q_{k_1\cdots k_d}^{(l_1\cdots l_d)}$ is nonzero if and only if $l_j - m_j + m_j'=0,~\forall j=1,\cdots, d$. Recall that $n_i = m_i- j_i > 0$, by changing $n_i$ to $m_i$ in Eq.~(\ref{FT_-1}) we eventually obtain the relationship between density matrix components and Fourier transforms of displaced fock state populations $Q_{k_1\cdots k_d}^{(l_1\cdots l_d)}$ (with summations truncated at $m'_{\textrm{max}}$):

\begin{align}
    Q_{k_1\cdots k_d}^{(l_1\cdots l_d)} 
    =& \frac{1}{(2N)^d}\sum_{p_1=-N}^{N-1}\cdots\sum_{p_d=-N}^{N-1}
    % \sum_{q=-N}^{N-1}
    \left[Q_{k_1\cdots k_d}(\alpha_{1,p_1},\cdots,\alpha_{d,p_d}) e^{-i\sum_{j=1}^{d}\left(l_j p_j\right)\pi/N}\right]\nonumber\\
    =& \sum_{m'_1=\mathrm{max}(0, -l_1)}^{m'_{\textrm{max}}}\cdots
    \sum_{m'_d=\mathrm{max}(0, -l_d)}^{m'_{\textrm{max}}}
    \gamma_{k_1m'_1}^{(l_1)}\cdots\gamma_{k_dm'_d}^{(l_d)}\rho_{l_1+m'_1,\cdots, l_d+m'_d, m'_1,\cdots, m'_d}
    % \label{FT}
\end{align}
% \end{widetext}
with 
\begin{align}
    \gamma_{k_im'_i}^{(l_i)} =& \frac{e^{-\left|\alpha_i\right|^2}\left|\alpha_i\right|^{2k_i}}{k_i!}
    \sum_{j_{i}=0}^{\mathrm{min}\left(k_i, m'_i+l_i\right)}
    \sum_{j_{i}'=0}^{\mathrm{min}\left(k_i, m'_i\right)}
    \left|\alpha_i\right|^{2(m'_i-j_i-j_i')+l_i}\cdot(-1)^{-j_{i}-j_{i}'}{k_i\choose j_i}{k_i'\choose j_i'}\frac{\sqrt{m'_i!(m'_i+l_i)!}}{{(m'_i-j_i')!(m'_i+l_i-j_i)!}}
    % \label{gamma}
\end{align}
and $i=1,\cdots, d$.

\end{document}